\documentclass{aa}  

\usepackage{txfonts}
\usepackage{graphicx,epsfig,fancyhdr,rotating,amsmath,natbib}
\def \kms {km~s$^{-1}$}
\def \rsun {R$_\odot$} 
\def \msun {M$_\odot$}

\def \vsi {$v \: \sin \: i$}

\begin{document}

   \title{Spectral observations of X Persei: Connection between  $H\alpha$ and X-ray emission}
   \titlerunning{X Per: Connection between  $H\alpha$ and X-ray emission} 
   \author{R. Zamanov\inst{1}\fnmsep\thanks{\email{rkz@astro.bas.bg,  kstoyanov@astro.bas.bg}}
          \and
          K. A. Stoyanov\inst{1}
          \and 
          U. Wolter\inst{2}  
          \and 
          D. Marchev\inst{3} 
          \and
          N. I. Petrov\inst{1}
          }
   \institute{Institute of Astronomy and National Astronomical Observatory, Bulgarian Academy of Sciences, 
                 Tsarigradsko Shose 72, BG-1784 Sofia, Bulgaria  
         \and
             Hamburger Sternwarte, Universit\"at Hamburg, Gojenbergsweg 112, 21029 Hamburg, Germany   
         \and
             Department of Physics and Astronomy, Shumen University, 115 Universitetska Str., 9700 Shumen, Bulgaria  
             }

   \date{Received November 1, 2018; accepted January 10, 2019}

 
  \abstract{We present spectroscopic observations of the Be/X-ray binary X Per obtained during the period 1999 - 2018.
  Using new and published data, we found that during "disc-rise" the expansion velocity of  
  the circumstellar disc is  0.4 - 0.7 \kms. Our results suggest that the disc radius in recent decades 
  show evidence of resonant truncation of the disc by resonances 10:1, 3:1, and 2:1, 
  while the maximum disc size is larger than the Roche lobe of the primary and 
  smaller than the closest approach of the neutron star. 
  We find correlation between equivalent width of $H\alpha$ emission line  ($W\alpha$)  and the X-ray flux, 
  which is visible when $15 \: \AA \:  < W\alpha \le 40$~\AA.
  The correlation is probably due to wind Roche lobe overflow.
  }


   \maketitle
%

\section{Introduction}
X Persei (HD 24534) is a relatively bright variable star, detected in X-rays with  $UHURU$ satellite 
and identified as the optical counterpart of the pulsating X-ray source 4U~0352+309 
\citep{1972Natur.235..273B, 1972Natur.235..273V}.
This object belongs to the class of Be/X-ray binaries which contains about 100 confirmed and suspected members
in our Galaxy, see e.g.   \citet{1994SSRv...69..255A} and   \citet{2011Ap&SS.332....1R}.
The main component of X~Per  is a hot massive rapidly rotating Be star. 
The secondary is a slowly spinning neutron star ($P_{spin} \approx  837$~s)
accreting matter from the wind of the  primary component \citep{2017MNRAS.470..713M}.
On the basis of data from Rossi X-Ray Timing Explorer, 
\citet{2001ApJ...546..455D} determined the orbital period $\sim$250~d,
orbital eccentricity $e$ = 0.11, and semi-major axis $a$ = 2.2~a.u. 
X~Per is considered prototype of the class of Be/X-ray binaries with long orbital periods and near-circular orbits  
\citep{2002ApJ...574..364P},  also named "low-e BeX" \citep{2004RMxAC..20...55N}.  
The low orbital eccentricity may be explained by a supernova explosion 
at which the neutron star did not receive a large impulse, or ``kick,'' at the time of formation, 
and  all induced eccentricity is due to mass loss.
\citet{2011Natur.479..372K} suggested that it may be related to the type of the  supernovae explosion which produced 
the neutron star, i.e. core-collapse or electron-capture supernova.

X~Per is a low-luminosity persistent X-ray source and  periodic X-ray outbursts are not obvious in its X-ray light curve. 
\citet{2007A&A...474..137L}  and \citet{2012MNRAS.423.1978L} detected X-ray flares and variability
in the X-ray light curve. \citet{2017MNRAS.470..713M} detected variability of the pulse profile and energy spectrum
and suggested that it is a result of changes in  the accretion geometry.
X~Per has an unusually hard X-ray spectrum \citep{1981ApJ...247L..31W}.
It is confidently detected by INTEGRAL up to more than 100 keV, which 
can be explained as a result of the dynamical Comptonization in the
accretion flow of photons emerging from the polar cap \citep{2012A&A...540L...1D}. 

In the last century, the visual magnitude of X~Per varies in the range  V= 6.8 to  6.2 mag.
The brightness variations are accompanied by variations in the intensity of the emission lines.
The  optical spectrum of X~Per shows strong emission in the Balmer lines, 
when it is brighter than 6.5 mag \citep{1979IAUC.3352....2D, 1998MNRAS.296..785T}. 
The brightness variability is probably connected with 
various disc-rise and disc-fade events.
During the last 50 years, the star exhibited two low states - probably disc-less episodes during 1974-1977 and  1990-1991. 
Absence of emission lines in 1977 was noted by \citet{1979A&A....78..287D}.   
\citet{1993A&A...270..122R} identified that the low state was  during 1974-1977. 
Dramatic changes were observed in  1990, when 
X~Per lost the   $H\alpha$ emission line,  infrared excess, and circumstellar disc \citep{1991MNRAS.253..579N}. 
The emission lines disappeared and the optical spectrum was dominated by absorption lines, 
which is typical for normal early-type star. 

We present optical spectroscopic observations obtained in recent decades 
and discuss the radius of the circumstellar disc, 
outflowing  velocity in the disc, disc truncation, 
and connection with X-ray emission.

 \begin{figure*}   
  \vspace{11.0cm}   
  \includegraphics{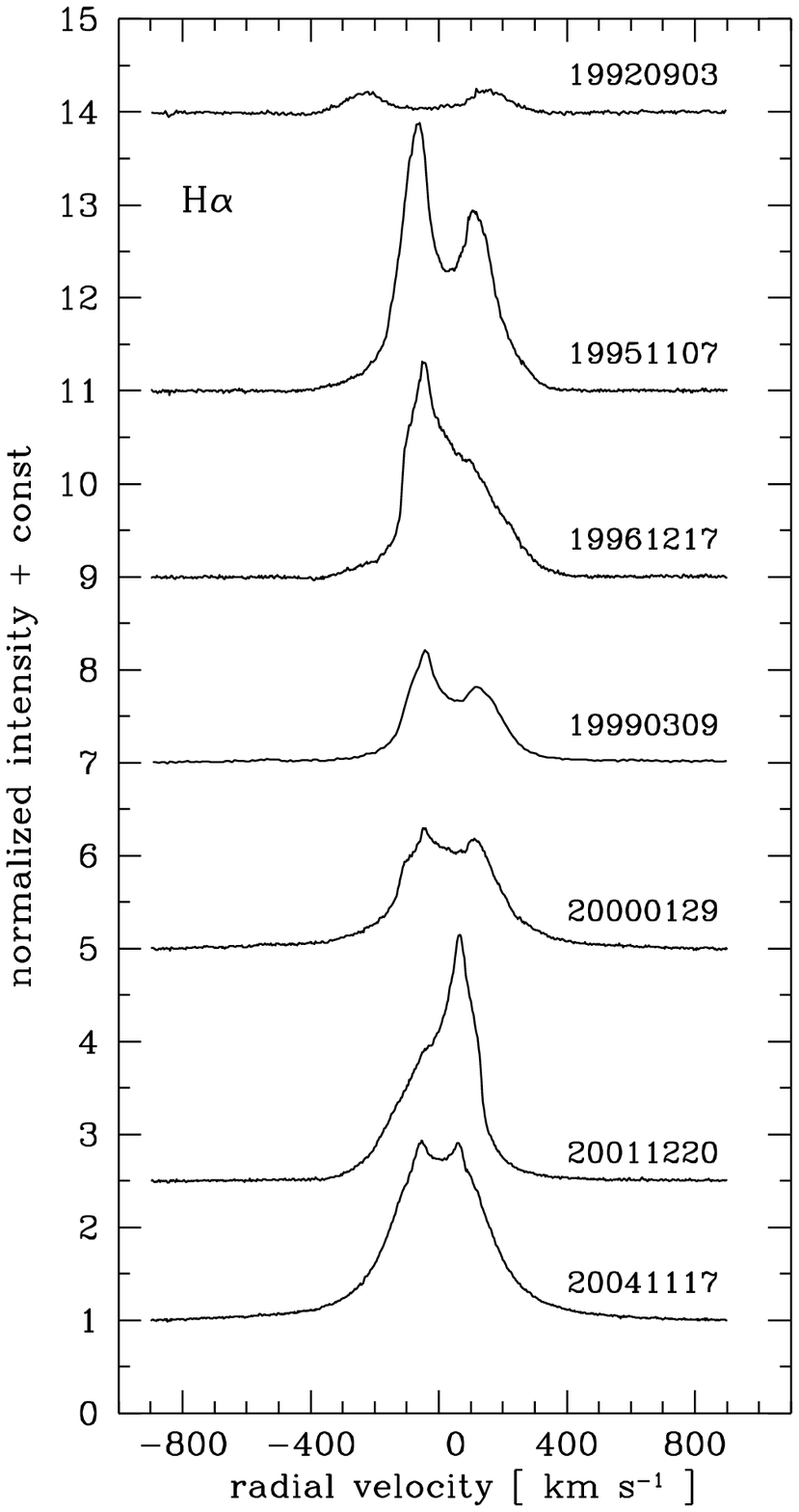}      
  \includegraphics{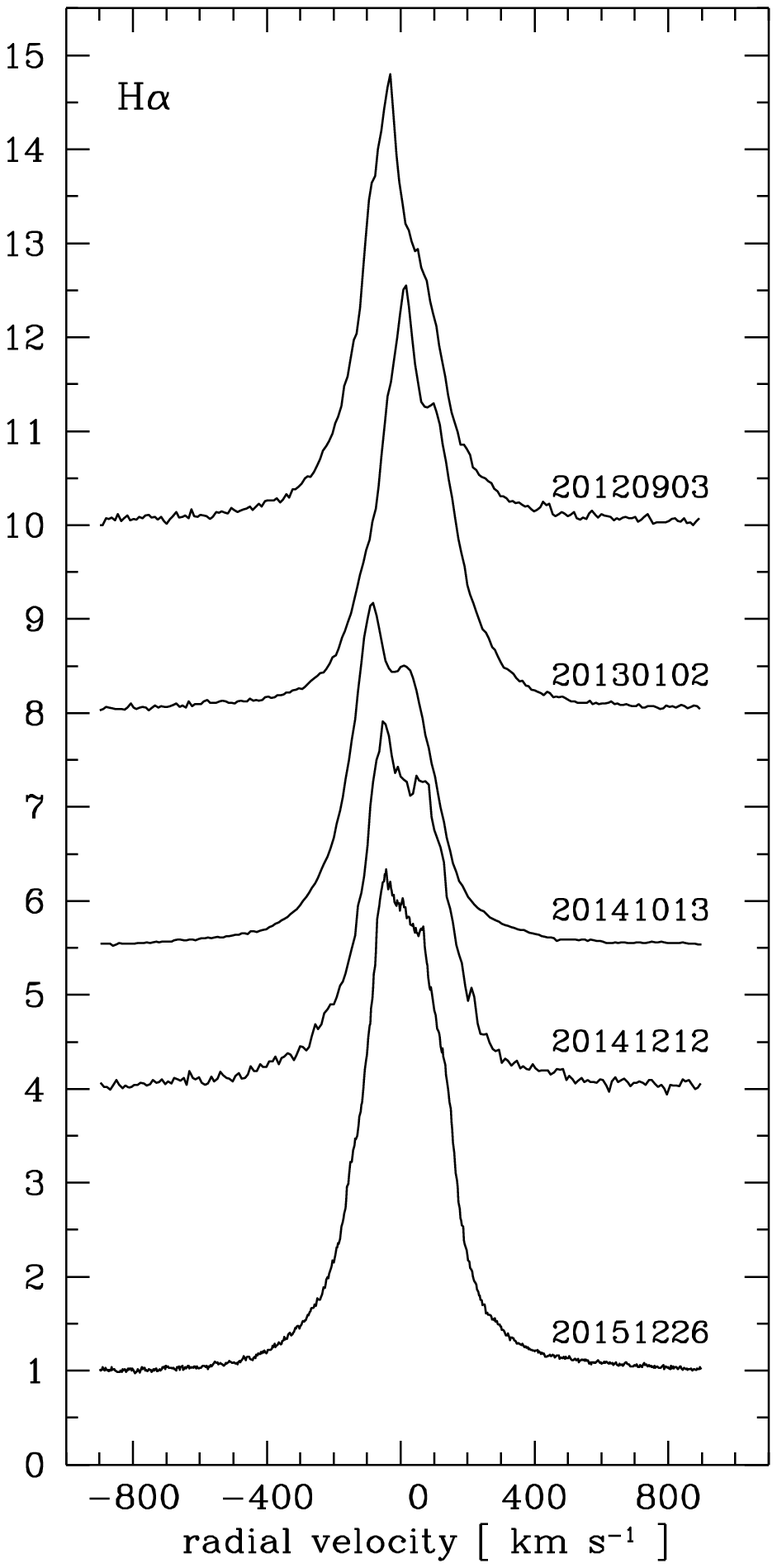}      
  \includegraphics{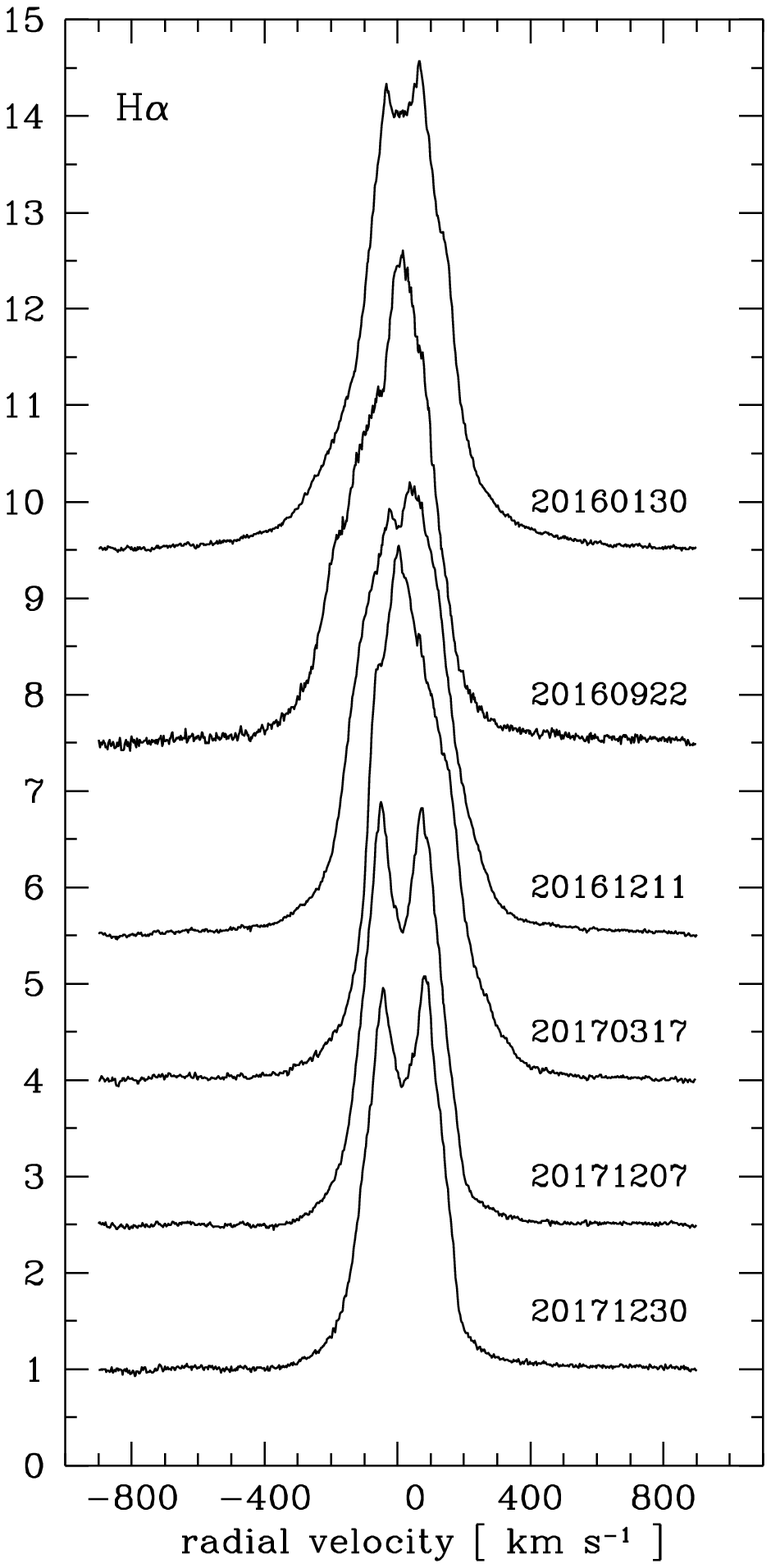}      
  \includegraphics{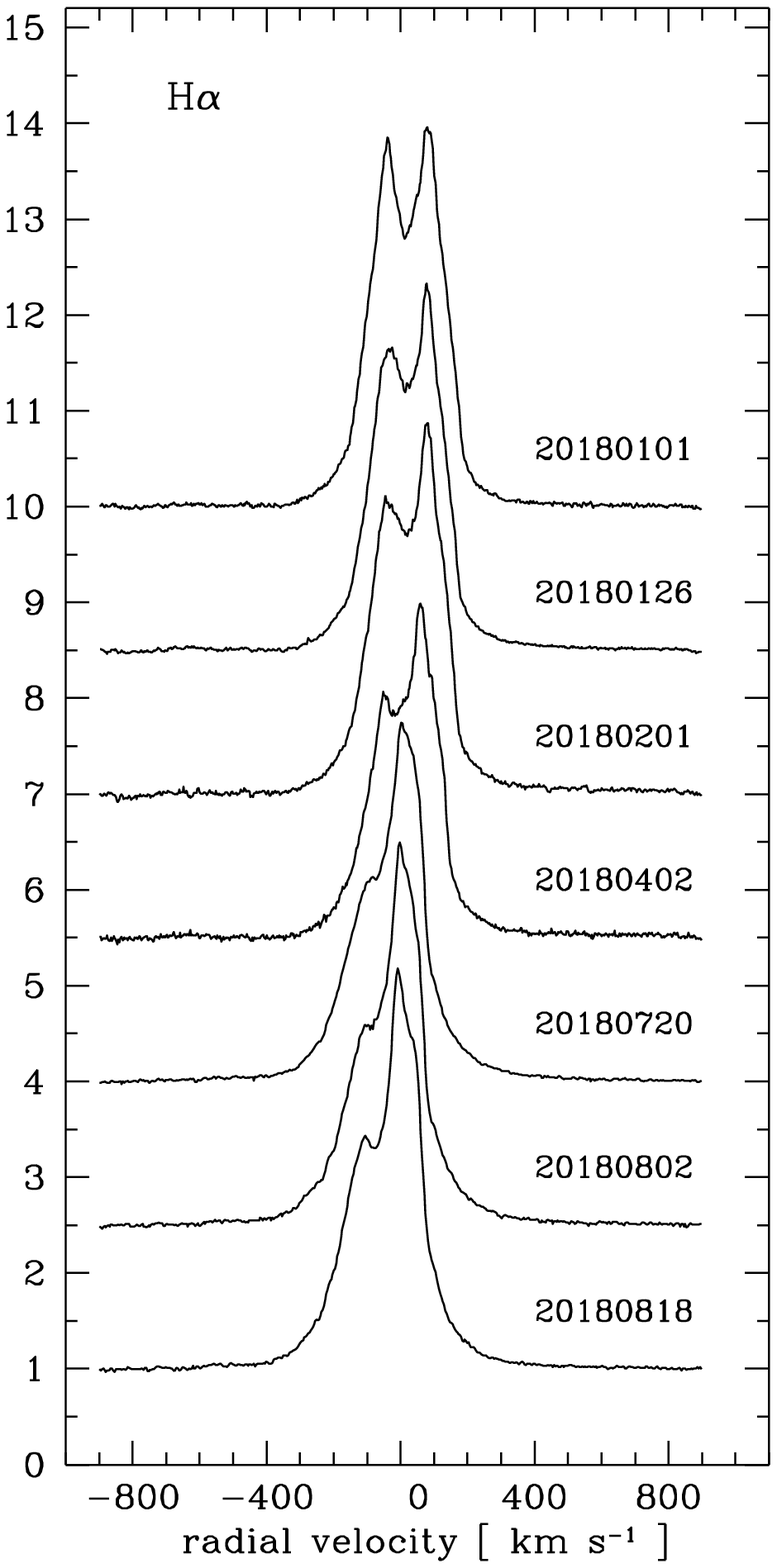}      
\caption[]{Variability of $H\alpha$ emission line profile of X Per.
  The date of observations is in format YYYYMMDD.
}  
\label{f1.atlas}      
\end{figure*}        

\section{Observations}

The optical spectra of X Per were secured with the 2.0 m telescope of the Rozhen National Astronomical Observatory, 
Bulgaria and with the robotic 1.2 m  TIGRE telescope located is the astronomical observatory La Luz in Mexico. 
The Rozhen spectra were obtained with the Coude spectrograph and with 
the  Echelle spectrograph ESpeRo \citep{2017BlgAJ..26...67B}.  
The Coude spectra have dispersion  0.1 \AA\  px$^{-1}$ or 0.2  \AA\ px$^{-1}$, while the Echelle spectra have  0.06 \AA\ px$^{-1}$ at 6560 \AA\ and 0.04 \AA\ px$^{-1}$  at 4800 \AA. 
The TIGRE spectra were obtained with  the HEROS spectrograph, which 
provides a spectral resolution of 20000 over the visual spectral range from 3800 \AA\ to 8800 \AA\
\citep{2014AN....335..787S}.
A few more spectra were downloaded from the ELODIE archive \citep{2004PASP..116..693M}.
These were obtained with the 1.93 m telescope of  Observatoire de Haute-Provence.

The measured parameters are given in Table~\ref{tab.obs}. 
The variability of  $H\alpha$ emission line of X Per is presented in Fig.\ref{f1.atlas} (emission line profile),
Fig.\ref{f2.EW.dV} (distance between the peaks), Fig.\ref{f3.vel} (comparison between two high resolution profiles), 
Fig.\ref{f.hist}  (histogram of the $H\alpha$ disc size), and Fig.\ref{f.HaX}  (long term behaviour). 
During the period 1992  - 2018, the equivalent of  $H\alpha$ emission line ($W\alpha$)
varies from 2~\AA\ up to  40~\AA. 
From Fig.\ref{f1.atlas}, it is visible that the  $H\alpha$ emission line, which exhibits  various profile shapes, in some cases 
is symmetric with two peaks that are well separated,
single-peak,  double-peak and  asymmetric profiles  
indicating  density inhomogeneities. 

\section{Primary component}
\label{primary}

\subsection{Radius and mass of the primary} 

From IR observations, \citet{2017ARep...61..983T} obtained 
temperature $T = 26000 \pm 1000$~K and radius  $R_{1} = 17.1 \pm 0.6$~\rsun\ for the donor star,  
adopting distance to the system $d$ = 1300~pc. 
GAIA Data Release 2  \citep{2016A&A...595A...1G,  2018A&A...616A...1G} 
gives parallax 1.234 mas, which corresponds to a distance  810 pc. 
With the GAIA distance the results of 
\citet{2017ARep...61..983T}  should give  $R_1 = 10.7$~R$_\odot$.

The  B-V and U-B colours of X Per in low state should represent the colours of the primary component. 
During the period  JD~2448545 (1991 October 15) - JD~2449046 (1993 February 27),
the brightness of X Per was low $V \approx 6.77$, and the circumstellar disc was very small.
At that time the average colours were  $(B-V)=0.09 \pm 0.03$  and 
$(U-B)=-0.68 \pm 0.03$ \citep{1995IBVS.4189....1Z}. 
Adopting interstellar extinction towards X~Per $E_{B-V}=0.356$ 
\citep{1982IAUS...98..423V, 
2017BlgAJ..27...10N}, 
we calculate dereddened colours 
$(B-V)_0 = -0.26 \pm 0.03$  and  $(U-B)_0 = -0.95 \pm 0.03$. Following \citet{1982lndf.book.....S}, 
these colours correspond to B1 III - V spectral type, which  is about one spectral type later than  B0 derived 
by   \citet{1997A&A...322..139R} and \citet{1997MNRAS.286..549L}, 
and O9.5III by  \citet{1992A&A...259..522F}. 

Using the well-known formula for the absolute V magnitude,   
we obtain $M_V=-3.86$. The bolometric magnitude is $M_{bol} = M_V + BC$, where  
the bolometric correction is $BC = -2.70$  \citep{1982lndf.book.....S, 2013A&A...550A..26N}. 
Using  the solar values  $T_\odot = 5780$~K  and $M_{bol}^{sun} =4^m.69$,
we calculate  $R_1 = 9.2$~R$_\odot$, which is similar to the above value
from \citet{2017ARep...61..983T}.


\citet{2010AN....331..349H}  give average  masses for B0V and B1V stars 15.0~\msun\ and 12.0~\msun, respectively. 
Hereafter for the primary  of X Per, we adopt 
radius     9.2~\rsun\ $ < R_1 < 10.7$~\rsun\ ($R_1 = 10.5 \pm 1.2$~\rsun) 
and mass 12.0~\msun\ $\le M_1 \le 15.0$~\msun\   ($M_1 = 13.5 \pm 1.5$~\msun).

\subsection{Rotation of the primary}  

\citet{1986A&A...159..276D} and \citet{1989Ap&SS.161...61H}  established a relation between  ${\rm v} \sin i$,
full width  at half maximum (FWHM), and $W\alpha$ of the $H\alpha$ emission line. To calculate ${\rm v} \sin i$, we use this relation in the form  
 \begin{equation}
    {\rm v} \sin i =  0.813 \; (FWHM \; 10^{0.08 \log W\alpha\ - 0.14} \; - 70 \; {\rm km \; s}^{-1})
    \label{Han1}
  ,\end{equation}
where FWHM and  $ \rm {v} \sin i$ are measured in \kms, W$_\alpha$ is in [\AA]. 
We measure FWHM and W$_\alpha$ on our spectra and obtain projected rotational velocity in the range 
$179 \le  \rm {v} \sin i \le 217 $~\kms 
with average $ \rm {v} \sin i = 191 \pm 12$~\kms.
This value is close to that of  \citet{1997MNRAS.286..549L}, who
estimated projected rotational velocity $v~sin~i$ = 215 $\pm$ 10~km~s$^{-1}$ using
HeI $\lambda 4026$ \AA\  absorption line.

 \begin{figure}   
  \vspace{11.0cm}   
  \includegraphics{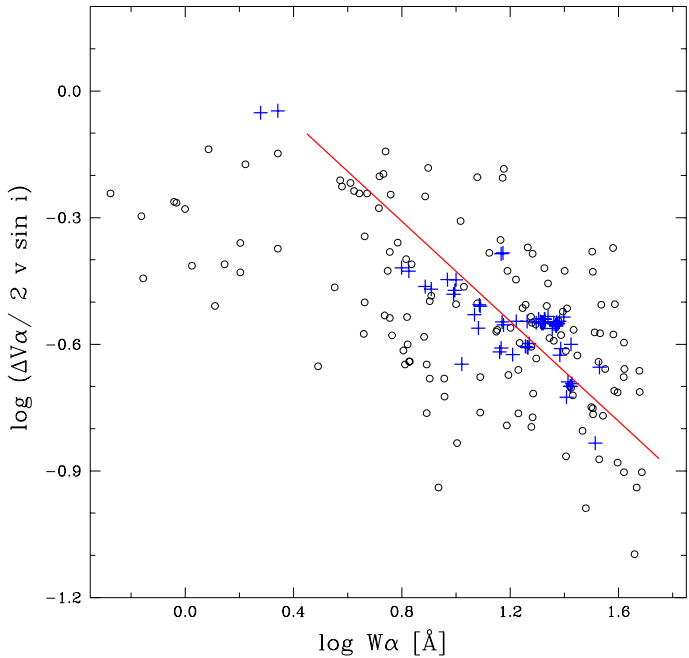}      
  \caption[]{Distance between the peaks normalized with stellar rotation  vs. $W\alpha$.  }  
  \label{f2.EW.dV}   
\end{figure}         

\section{Be disc}
\label{Disc.size}

In Fig.~\ref{f2.EW.dV}, we plot $\log \Delta V \alpha / 2 $\vsi \   versus $\log W\alpha$. 
The black open circles are data for Be stars taken from 
\citet{1983A&AS...53..319A},   
\citet{1986A&A...166..185H},   
\citet{1988A&A...189..147H},    
\citet{1992A&AS...95..437D},    
\citet{1992ApJS...81..335S},    
and \citet{2013A&A...550A..79C}.    
The blue plus signs are our measurements of X~Per. Most of the data points of X Per are in the middle of the 
Be star populations. The exceptions are two points obtained on 19920903  and  19920905, when the disc was small.
Because these points are above the population of other Be stars, following 
Sect.~5.2 of Hanuschik et al. (1988), this probably indicates that the disc of X Per was denser  
during the early stages of its development. Later, when the disc is larger, it follows the average behaviour of other Be stars.
For X~Per we have 15 spectra on which both $H\alpha$ and $H\beta$ emission lines have two peaks (see Table~\ref{tab.obs}).
We calculate median value $\Delta V_\beta / \Delta V_\alpha = 1.24$
and average value $\Delta V_\beta / \Delta V_\alpha = 1.30 \pm 0.26$. 


The Balmer emission lines form primarily in the disc surrounding the Be star, and the total flux of the feature (measured as 
the line equivalent width) is closely related to the size of the disc. 
The discs of the Be stars are Keplerian supported by the rotation (e.g. 
\cite{2003PASP..115.1153P}  
and references therein). 
For a Keplerian circumstellar disc the peak separation can be regarded as a measure of 
the outer radius ($R_{disc}$) of the emitting disc,  
 \begin{equation}
  R_{disc} = R_1 \frac{ (2\,v\,\sin{i})^2} {\Delta V ^2 }, 
  \label{Huang}
  \end{equation}
where  $R_1$ is the radius of the primary, $v\,\sin{i}$ is its projected rotational velocity. 
When the two peaks are visible in the emission lines, we can estimate the disc radius using Eq.~\ref{Huang}.
The disc size is also connected with  $W_\alpha$,
 \begin{equation}
       R_{disc}  = \; \epsilon  \; R_1 \; 0.467 \; \; W_\alpha^{1.184}, 
  \label{Rd.W2}
  ,\end{equation}
where $\epsilon$ is a dimensionless parameter, for which  we adopt $\epsilon = 0.9 \pm 0.1$
(see \citet{2016A&A...593A..97Z}, Sect. 4.3).  This equation 
expresses the fact that $R_{disc}$ grows as $W_\alpha$ becomes larger 
\citep[e.g.][]{2006ApJ...651L..53G}. 
A slightly different expression for the relation between  $R_{disc}$ and $W_\alpha$
is used in \citet{2006MNRAS.368..447C}  and \citet{2017MNRAS.464..572M}.

 \begin{figure}   
  \vspace{7.0cm}   
  \includegraphics{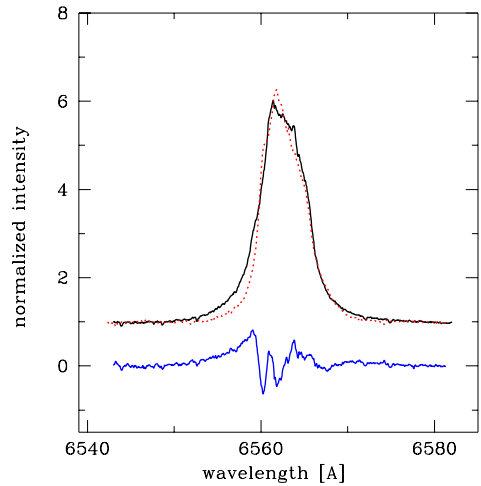}      
\caption[]{$H\alpha$ emission line profile of X Per in December 2015 (black solid line) and  March 2017 (red dotted line). 
   The two are spectra obtained 
  about one year apart.  The difference between the spectra is also plotted and seems to 
  indicates slow outward motion with velocity $\approx 5.1$  km~s$^{-1}$ 
  (see Sect.~\ref{vout}). }  
\label{f3.vel}      
\end{figure}         

\subsection{Radial outflow velocity}

\label{vout}

During the period JD~2448869 to JD~2450029 a steady increase of the H$\alpha$  emission is visible (Fig.\ref{f.HaX}).  
The equivalent width increased from  W$\alpha = -2.0$~\AA\ 
to  -14.9~\AA,   
and the distance between the peaks decreased
from $\Delta V = 382$ \kms\ to  $\Delta V = 177$ \kms.
Following Eq.~\ref{Huang} this corresponds to change of the disc radius from  13~\rsun\  to  62~\rsun\ and 
expansion velocity of disc  $V_{out} = 0.35$~\kms.
Using W$\alpha$ and  Eq.~\ref{Rd.W2} this corresponds to change of the disc 
radius from  10~\rsun\  to  108~\rsun\ and 
expansion velocity of disc  $V_{out} = 0.70$~\kms.

During the period JD~2452800 to JD~2455500 a steady increase of the H$\alpha$  
emission is also visible (Fig.\ref{f.HaX}).
The equivalent width increased from  W$\alpha = -12$~\AA\ to  -37~\AA.
Using  Eq.~\ref{Rd.W2}, this corresponds to $V_{out} = 0.67$~\kms.

We compared the high resolution emission line profiles obtained with the ESPERO spectrograph. 
An interesting result emerged when we compared the  $H\alpha$  profiles obtained in  December 2015
with that in March 2017 (see Fig.~\ref{f3.vel}). 
It seems that during this period the material that was emitting in the wings of the $H\alpha$ emission
moved to the outer parts of the disc and on the later spectrum this material emits in the central part of the line.
Ring-like structures in Be discs are discussed for example by 
\citet{1931ApJ....73...94S}  and \citet{2001A&A...379..257R}.    
In X~Per,  \cite{1995MNRAS.276L..19T} detected the appearance and development of an inner ring-like structure in the period
1993-1995, when the disc was starting to rebuilt after a disc-less phase.

Supposing that the variability seen on Fig.~\ref{f3.vel} 
represents a ring-like structure (a ring-like density enhancement) in the disc, we  calculate 
that the ring initially emits at $\Delta V = 220$~\kms,  and later at  $\Delta V = 78$~\kms. 
Applying  Eq.\ref{Huang} we obtain that the material moved from 40~\rsun\ to 325~\rsun\  for 447 days, having average 
outflow velocity 5.0~\kms. This velocity is about eight times faster than the velocity
we estimated from disc build-up, and might be an indication that once the disc is developed the material can 
move faster inside the disc.

 \begin{figure}   
  \vspace{16.0cm}   
  \includegraphics{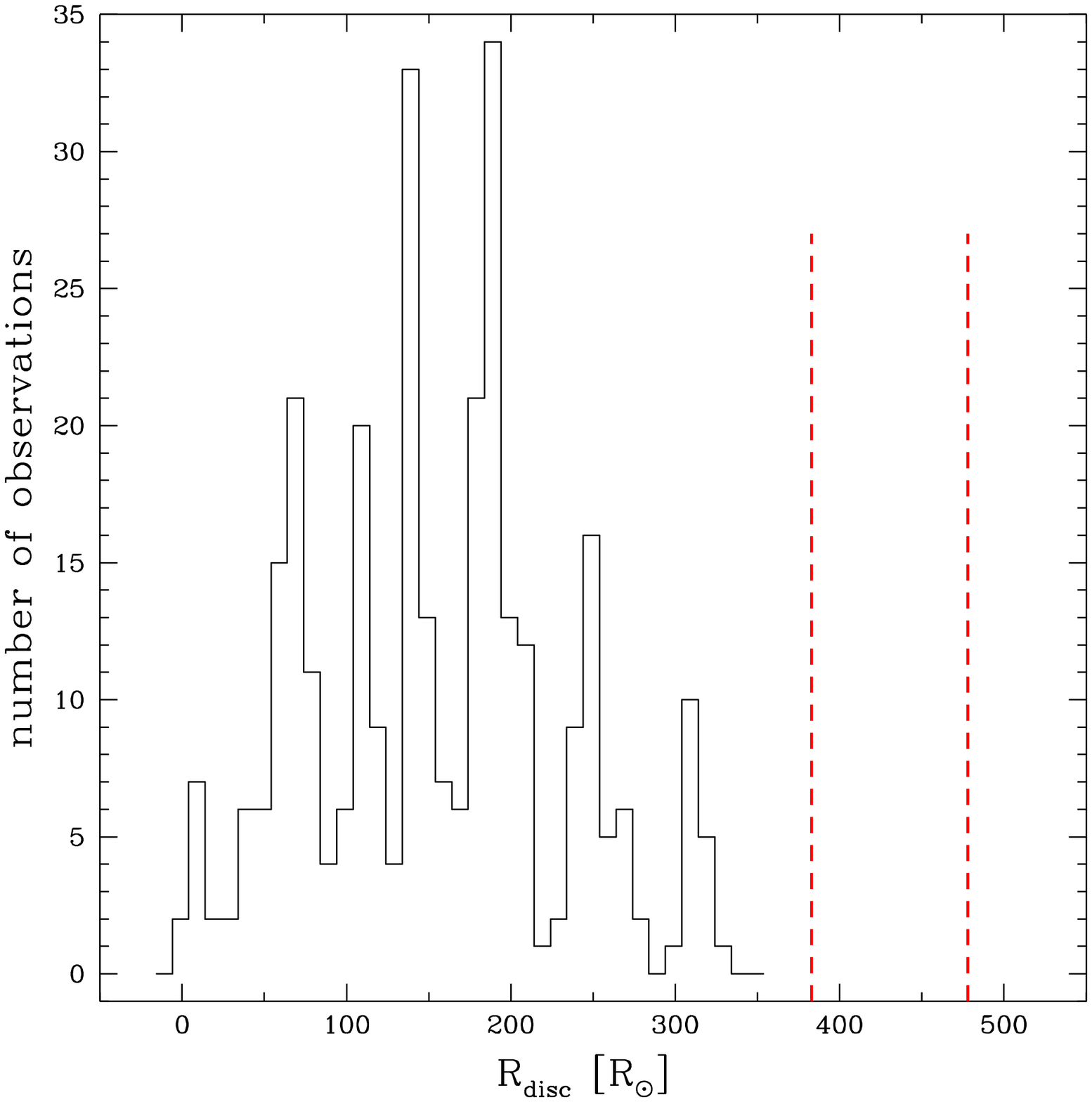}      
  \includegraphics{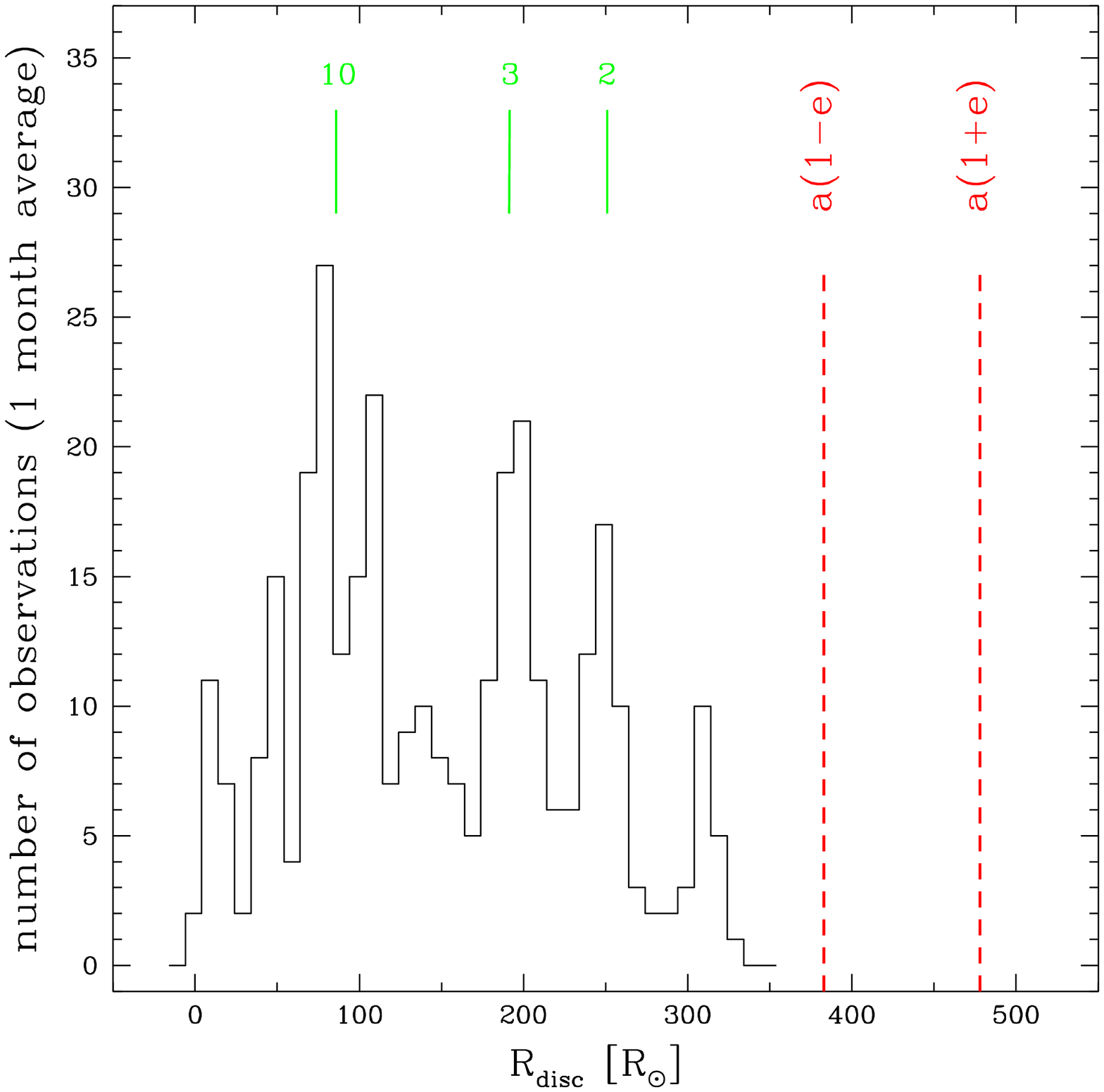}      
\caption[]{Distribution of the calculated disc radius. The red dashed lines indicate the distance between 
components at periastron and apastron; the green solid lines show the resonances. 
The upper panel show the observed values; the lower indicate 30 days binned.}  
\label{f.hist}      
\end{figure}         

\subsection{Disc truncation}
\label{disc.trunc}


The tendency for the disc emission fluxes to cluster at specified levels is
related to the truncation of the disc at specific disc radii by the orbiting compact object
\citep[e.g.][]{2006MNRAS.368..447C}. 
\citet{2001A&A...377..161O} proposed
that these limiting radii are defined by the closest approach of the companion
in the high-eccentricity systems and by resonances between the orbital period and disc gas rotational periods in 
the low-eccentricity systems. The resonance radii are given by
 \begin{equation} 
    {R_{n:m}^{3/2}} = \frac{m \; (G \: M_1)^{1/2}}{2 \: \pi} \:  \frac{P_{orb}}{n}, 
  \label{eq.resona}
  \end{equation} 
where $G$ is the gravitational constant, $n$ is the integer number of disc gas rotational periods, and
$m$ is the integer number of orbital periods. 
The important resonances are not  only those with $n:1$, but can as well be $n:m$ in general.

The histograms of $H\alpha$ disc size,
$R_{disc}$, as  calculated by Eq.~\ref{Rd.W2}, are plotted in Fig. 4.  
We use data for W$\alpha$ from  
\citet{2016A&A...590A.122R},    
\citet{2007ApJ...660.1398G}, and    
\citet{2014AJ....148..113L},    
as well as  data from Table~\ref{tab.obs}. 
Fig.~\ref{f.hist} (the upper panel) is the histogram of the measured values of  $R_{disc}$.
This histogram is affected by the distribution of the observations, for example 
if we have many observations in one month it could produce a spurious peak in the histogram. 
To correct for this effect we calculate the average values for each 30 day. 
If there is no data for a given interval, we interpolate between the nearest values.
The 30-day binned values are plotted in Fig. 4 (lower panel). 
The 30-day bin was used because we do not detect variability in $H\alpha$ on a timescale 
shorter than 10 days, but we do detect  variability on 60 days. Binning with  20 and 40 days provide 
very similar results.
The comparison between Fig.~\ref{f.hist}a and  Fig.~\ref{f.hist}b shows that the spurious effects due 
to unequal distribution of the observations are removed.

Fig.~\ref{f.hist} is based on data over the last 26 years from 1992 to 2018. 
We note that in three X-ray/$\gamma$-ray binaries  (LSI+61~303, MWC 148, and MWC 656)
the distribution of $R_{disc}$ values has one very well-pronounced peak \citep{2016A&A...593A..97Z}.  
In  X Per we see a few peaks rather than a single peak.  
The most pronounced peaks correspond to 10:1, 3:1, and 2:1 resonances.  
In X~Per it seems that in the beginning of the disc-rise the resonance 10:1  operates, 
after this more disc material appears, the disc 
grows, and the resonance 3:1  starts 
to operate, and later  2:1. 
A more or less similar situation is in the Be/X-ray binary 
V725~Tau~/~1A~0535+262, where multiple resonances are discussed, i.e. 1:4, 1:5, and 1:7 \citep{2006MNRAS.368..447C}.  
The orbital period of  V725~Tau is 111.1~days  \citep{1996ApJ...459..288F}.      
This similarity  indicates that in such wide systems different resonances can operate 
probably     
depending on the mass loss of the primary and the development of its circumstellar disc. 

Following  \cite{2018PASJ...70...89Y},   
we adopt  $P_{orb}=251.0 \pm 0.2 $~d and mass of the neutron star $M_{ns}=2.03$~\msun.
For the primary we assume $M_1=13.5$~\msun\  (see Sect.~\ref{primary}). 
With these values we calculate mass ratio 
$q=M_1/M_{ns} = 6.65$ and semi-major axis of the orbit  $a=418$~\rsun.     
Using the formula by \citet{1983ApJ...268..368E},  
we estimate the Roche lobe size of the primary $r_L = 228$~\rsun.
The orbital eccentricity of the system is low $e \approx 0.11 $ 
\citep{2001ApJ...546..455D}.     
The distance between components at periastron is $ a \:(1-e) = 372$~\rsun\  and
at  apastron is $ a \: (1+e) = 464$~\rsun. 

The maximum disc size observed in our data is  $R_{disc} = 337$~\rsun, which is smaller than the 
closest approach of the neutron star.  However, it is larger than the size of the Roche lobe of the Be star, i.e. 
$r_L   <  R_{disc} (max) <  a (1-e)$.

The Be/X-ray binaries  present  different states of X-ray  activity 
\citep{1986ApJ...308..669S, 1998A&A...338..505N}.                   
X~Per does not show periodic Type I outbursts, which are common feature in the classical Be/X-ray binaries
and occur when a neutron star moving along an eccentric orbit crosses the Be circumstellar disc 
during the periastron passage. 
The result that  $R_{disc} (max) <  a (1-e)$ 
demonstrates  that the neutron star does  not pass through the disc of the donor star even at the maximum 
disc size observed during the last 30 years. Periodic Type I outbursts in X~Per can be expected 
if $W\alpha$  achieves  the level of about 45 \AA\ or above it.   


 \begin{figure}   
  \vspace{15.0cm}   
  \includegraphics{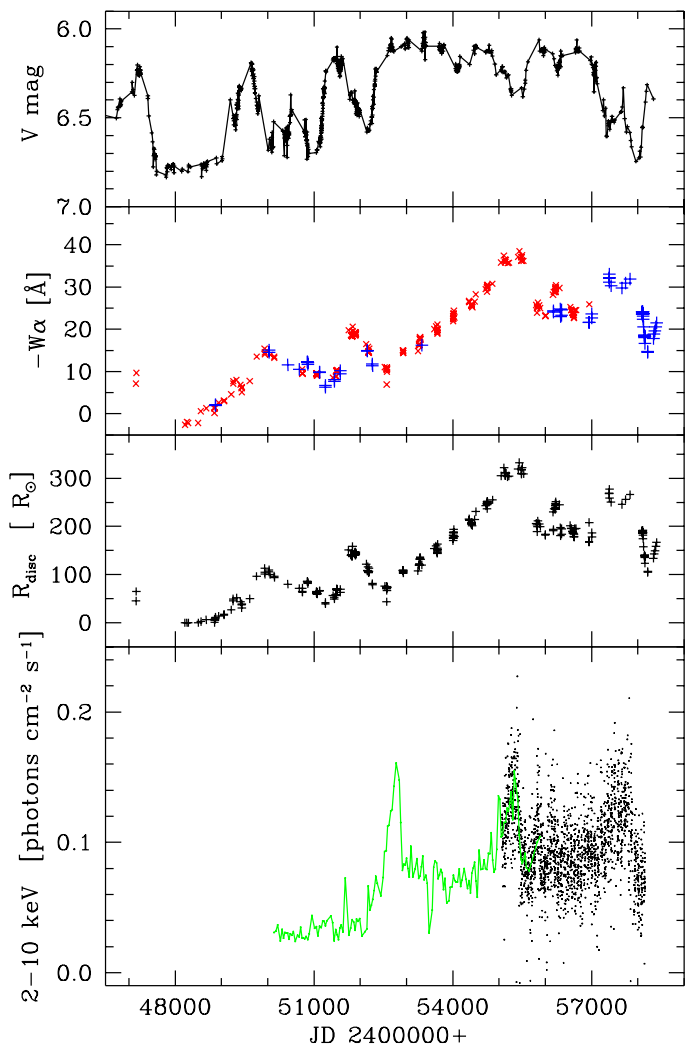}      
  \caption[]{X Per - connection between $H\alpha$ and X-ray variability.  
Shown from top to bottom are V-band magnitude,
equivalent width of $H\alpha$ emission, circumstellar disc radius,  and X-ray flux. }
\label{f.HaX}      
\end{figure}         

\subsection{Be disc -->  X-ray flux }


The mass transfer rate onto the neutron star  and subsequent accretion-driven X-ray flux 
should depend on the changes in the radius of the disc of the Be star that we can track through the variations in H$\alpha$ emission line.  
The  V-band magnitude, 
$W\alpha$, the calculated disc radius, and X-ray data are plotted in Fig. 5. 
The V-band data are from  American Association of Variable Star Observers (AAVSO).
The X-ray data are from  RXTE/ASM  
\citep{1996SPIE.2808...59J, 1996ApJ...469L..33L} 
and MAXI   \citep{2009PASJ...61..999M}.
The RXTE/ASM X-ray light curve 
is rebinned 
with a time of 30 days and scaled relatively to the MAXI counts 
(binning with  25 or 35 days provide very similar light curve). 
In the panel  $W\alpha$,   the red crosses are data from the literature 
\citep{2016A&A...590A.122R, 2007ApJ...660.1398G, 2014AJ....148..113L}.
The blue plus signs represent Rozhen, TIGRE, and ELODIE data from our Tables~\ref{tab.obs} and~2 of \cite{2001A&A...367..884Z}.

\cite{2014AJ....148..113L}   
discussed the connection between $H\alpha$  and X-ray flux 
during the period JD~2451000 - JD~2455000. They pointed out the time delay 
between the   $W\alpha$   maximum at about JD~2452000 and X-ray maximum, which is $\sim 800$ days later.
They used this delay to estimate the viscosity in the outflowing disc ($\alpha$ parameter). 
We point that there is a similar time delay between  a minimum in  $W\alpha$  
at JD~2452600 and a minimum in RXTE/ASM X-ray flux at JD~2453500. 
 
After JD~2454000  $W \alpha $ increases and the X-ray flux also increases, which indicates that
the neutron star accretes more material and probably  truncates the outer parts of the disc.
Two stronger peaks are visible in $W\alpha$ at JD~2455000 and JD~2457500. Both of these peaks are also 
well detectable in the MAXI X-ray data. 
We performed correlation analysis and found moderate to strong
correlation between $W\alpha$ and X-ray flux
with the Pearson correlation coefficient 0.61-0.66, 
Spearman's (rho) correlation coefficient 0.61-0.68, 
and significance, $p-value$ in the range $5 \times 10^{-6}$ - $1 \times 10^{-8}$.
There is no  time delay between $W\alpha$ and X-ray flux 
after JD~2454000. If a time delay exists  it is less than 40 days.

A comparison between the behaviour of V brightness and $W\alpha$ shows that before 
JD~2453000 these parameters
vary almost simultaneously -- the maxima of 
$W \alpha $  correspond to the maxima of the V-band  brightness and have  
time delays $\lesssim 300$ days.
After it, the behaviour is different  
-- the maxima of $W \alpha $  and $R_{disc}$ correspond to minima of the V-band brightness. 

%
%




\section{ Discussion }
\label{Discussion}

 For Be stars, \cite{1988A&A...189..147H} 
found that  the peak separations of $H\alpha$ and $H\beta$ emission 
lines follow the relation $\Delta V_\beta = 1.8 \Delta V_\alpha$. 
For X~Per the ratio $\Delta V_\beta / \Delta V_\alpha$ is 
considerably below the average value for Be stars (see Sect. \ref{Disc.size}). 
At this stage considerable deviation from the behaviour of the Be stars is also detected in LSI$+61^0303$
\citep{2016A&A...593A..97Z}.           
In both star the $H\alpha$-emitting disc is only 1.7 times larger than 
the $H\beta$-emitting disc, while in normal Be 
stars it is 3.2 times larger. This can be a result of  truncation
of the outer parts or  different density structures in the inner parts of the Be disc. 


The pulse period of neutron star in X~Per shows episodic spin-ups and spin-downs. 
Between 1972 and  1978 the neutron star has been spinning 
up with a rate of $\dot{P}/P \simeq -1.5 \times 10^{-4}$~yr$^{-1}$ \citep{1983adsx.conf..393H}. 
Since 1978 till 2002 the neutron star has been spinning down with a rate 
$\dot{P}/P \simeq 1.3 \times 10^{-4}$~yr$^{-1}$. 
After 2002 (JD~2452000) a new episode of spin-up has began
\citep{2012int..workE..26B,    2014MNRAS.444..457A}. 
This new spin-up began together with the increase of the X-ray brightness, and its start corresponds to the moment when 
$W\alpha$ achieved 20~\AA\  and $R_{disc} \approx 150$~\rsun.  

\cite{2018PASJ...70...89Y} 
analysed X-ray data and estimated for the neutron star
in X Per mass $M_{ns} = 2.03 \pm 0.17$ \msun,
and magnetic field  $B = (4 - 25) \times 10^{13}$~G, adopting GAIA distance $d = 810$~pc. 
They also found that the  X-ray luminosity varies in the 
range $2.5 \times 10^{34}$ to $1.2 \times 10^{35}$ erg s$^{-1}$. 

\cite{2017MNRAS.465L.119P}     
proposed that  in the enigmatic Be star $\gamma$~Cassiopeia
the elusive companion is a  neutron star acting as a propeller. 
In this scenario, the subsequent evolutionary stage of $\gamma$~Cas and its analogues should be X Per-type binaries
comprising low-luminosity slowly rotating X-ray pulsars.
The corotation radius of the neutron star is
\begin{equation}
R_{co} =  \left( \frac{G \; M_{ns} \; P_{spin} ^2}{4 \;  \pi ^2} \right) ^{1/3}
.\end{equation}
The radius of its magnetosphere (the Alfv\'en radius) is 
\begin{equation}
R_{m} =  \frac{ \left( B \; R_{ns}^3  \right) ^{4/7}}{ \dot M_{acc}^{2/7} \; \left( 2 \; G \;  M_{ns} \right) ^{1/7} }
  \label{eq.Rm}
.\end{equation}
In the standard theory of gravimagnetic rotators 
\citep{1987ans..book.....L,  2018A&A...610A..46C}    
for a neutron star to be accretor (X-ray pulsar)
the condition is that the magnetosphere radius should be
smaller than the corotation radius, $R_{m} \le R_{co}$.
Assuming $R_{ns}=10$~km  and  that X-ray flux is equal to the accretion luminosity, 
\begin{equation} 
    L_{acc} = G M_{ns} \dot M_{acc}  R_{ns}^{-1}, 
\label{Eq.a1}
\end{equation}
we  estimate  that the mass accretion rate is in the range $1.47 \times 10^{-12}$  to $7.07 \times 10^{-12}$  \msun\ yr$^{-1}$
(from $9.28 \times 10^{13}$  to  $4.45 \times 10^{14}$~g~s$^{-1}$).
We calculate that at low accretion rate  $R_{m} \approx R_{co} \approx 0.24$~\rsun, in other words 
at low X-ray flux the neutron star is close to the accretor-propeller transition boundary. 
The condition that the neutron star in X~Per is an accretor (not propeller), $R_{m} \le R_{co}$, 
also puts a limit on the magnetic field strength, $B \le 1.15 \times 10^{14}$~G. 
This value is estimated taking into account the deviation of the magnetosphere radius from 
the the Alfv\'en radius  \citep[e.g. ][]{2018A&A...617A.126B}. 
If the magnetic field is above this value, the neutron star would act as a propeller, 
the X-ray luminosity would decrease and the X-ray pulsations 
would disappear (which is not observed). 


\cite{1993A&A...270..122R}   
found a clear correlation between the optical,
infrared, and X-ray behaviours during the 1974-1977 low state, 
followed by an extended period in which the X-ray behaviour 
appears to be unrelated to the optical. 
Li et al. (2014), analysing data obtained after the disc-less episode in 1990, 
found a time delay of about 800 days between   $W\alpha$  and X-ray flux.
This time delay exists before JD~2454000. 
After JD~2454000 (as visible on Fig.~\ref{f.HaX}) there is no time delay, 
but there is a correlation between the variability of $W\alpha$ 
and X-ray flux. As the $W\alpha$ increases, the X-ray flux also increases.  
This is indicating a direct linkage between the circumstellar disc
around mass donor and accretion rate onto the neutron star. 
This linkage is visible when  $W_\alpha > 20$ \AA\ and $R_{disc} > 200$~\rsun. 
When the circumstellar disc around the primary increases
above 200~\rsun, the quantity of the  material that is captured in the accretion cylinder of the neutron star also increases.
The Roche lobe size of the primary is  $r_L = 227$~\rsun\ (see Sect.~\ref{disc.trunc}). 
When $W_\alpha \approx 25$~\AA,   $R_{disc} \approx r_L$, the  circumstellar disc fills the Roche lobe around the primary
and the neutron star begins to accrete material from the Roche lobe overflow. 
In the past few years the  existence of a relatively efficient mode of wind mass 
accretion in a binary system has been proposed, called the wind Roche lobe overflow 
\citep[WRLOF; ][]{2007ASPC..372..397M},      
which lies in between the canonical  Bondi-Hoyle-Littleton accretion and Roche lobe overflow.
In case of X~Per, most probably before JD~2454000 the circumstellar disc is small, it is well inside 
the Roche lobe ($R_{disc} < r_L$)
and wind accretion (Bondi-Hoyle-Littleton  accretion)  is acting. 
After JD~2454000 when the circumstellar disc size achieves the Roche lobe,  $R_{disc} \gtrsim r_L$,  
the accretion mode changes to WRLOF, 
and the neutron star is accreting material directly from the circumstellar disc 
(Roche lobe overflow from the circumstellar disc) without time delay. 
 
Other possibilities include the development of spiral arms \citep{2007ApJ...660.1398G} 
or other large-scale perturbations  \citep{1998A&A...336..251N}  
in the circumstellar envelope 
excited by tidal interaction which may lift disc gas out to radii where the accretion 
by the neutron star  becomes more effective. 




\section{Conclusions}
We present optical spectroscopic observations of the Be/X-ray binary X~Per, 
optical counterpart of the X-ray pulsar  4U~0352+309. 
In this work, we combine published data with our measurements.
First, we estimate, that the expansion velocity of the circumstellar disc is in the range 0.4 - 0.7 \kms .
Second, we find that the distribution of the disc radius from the average equivalent width 
          suggests resonant truncation of the disc, while the maximum disc radius
          is smaller than the separation of the stars at periastron but larger than 
          the Roche lobe of the primary. 
Third, we derive a correlation  between the equivalent width of $H\alpha$ emission line 
          and  X-ray flux, which is visible since JD~2454000, when $15~\AA \: < W\alpha \le 40$ \AA.
We briefly discuss possible mechanisms of mass transfer.

\vskip 1.0cm

\begin{acknowledgements} This work is supported by the grant K$\Pi$-06-H28/2 08.12.2018 
(Bulgarian National Science Fund). 
It is based on observations from Rozhen National Astronomical Observatory, Bulgaria and 
the TIGRE telescope, located at La Luz, Mexico. 
TIGRE is a joint collaboration of the Hamburger Sternwarte, the University of Guanajuato 
and the University of Li\`ege.
This research has made use of (1) the MAXI data provided by RIKEN, JAXA, and the MAXI team; 
(2) results provided by the ASM/RXTE teams at MIT and at the RXTE SOF and GOF at NASA's GSFC; and 
(3) observations from the AAVSO International Database 
contributed by observers worldwide.
UW acknowledges funding by DLR, project 50OR1701.
DM acknowledges partial support by grants  DN~08/20/2017 and  RD-08-112/2018.
We are very grateful to the referee whose comments helped to improve considerably the original
manuscript.

\end{acknowledgements}

%
%

\bibliographystyle{aa}
\bibliography{ref2.bib}


\newpage 

\begin{table*}
  \centering
  \caption{Spectral observations of X Per. 
   The spectra marked with (*) are partly published in \cite{2001A&A...367..884Z}.
}
  \begin{tabular}{lclrr ccr ccc  cccccr} 
\hline
file                         &   telescope   &  JD-start   &   exp-time  & W$_\alpha$  & $\Delta V\alpha$ &  $\Delta V\beta$ & \\
                             & instrument    &  2400000+   &     [min]   &   [\AA]      &   [\kms]         &   [\kms]         & \\
\hline
19920903*         & 2.0m Coude & 48869.5278 &  59   & -1.9 &       382 &    \\
19920905*          & 2.0m Coude & 48871.5042 &  54   & -2.2 &       386 &    \\
19951107.0024     &  Elodie       &  50029.4808  &  30 & -14.9  & 177.8   &  202.0 & \\
19951107.0028     &  Elodie       &  50029.5449  &  45 & -14.7  & 176.9   &  200.0 & \\
19961217.0024     &  Elodie       &  50435.3818  &   8 & -12.1  &         &         & \\
19970817*         & 2.0m Coude & 50677.5447 &  15   & -10.5 &        97 &    \\
19980209a*        & 2.0m Coude & 50854.1816 &   5   & -12.3 &       133 &    \\
19980209b*        & 2.0m Coude & 50854.1856 &   5   & -12.2 &       134 &    \\
19980219a*        & 2.0m Coude & 50864.2089 &   7   & -12.1 &       118 &    \\
19980219b*        & 2.0m Coude & 50864.2191 &   7   & -11.7 &       127 &    \\
19981102a*        & 2.0m Coude & 51120.6553 &   15  &  -9.8 &       142 &    \\
19981102b*        & 2.0m Coude & 51120.6660 &   15  &  -9.9 &       145 &    \\
19981230.65       & 2.0m Coude & 51178.4811 & 25 &  -8.4 & 167.7    \\
19981230.66       & 2.0m Coude & 51178.4991 & 25 &  -8.2 & 164.8    \\
19990309a*        & 2.0m Coude & 51247.2356 & 20 &  -6.7 & 161 &    \\
19990309b*        & 2.0m Coude & 51247.2497 & 20 &  -6.3 & 164 &    \\
19990919a*        & 2.0m Coude & 51441.5191 & 20 &  -8.1 & 146 &    \\
19990919b*        & 2.0m Coude & 51441.5333 & 10 &  -7.7 & 148 &    \\
20000129.0005     &  Elodie       &  51573.2545  &  30 &  -9.3  & 153.7   &  190.3 & \\
20000129.0006     &  Elodie       &  51573.2767  &  30 & -10.0  & 153.6   &  193.5 & \\
20010903.178      &  2.0m Coude   &  52156.5933  &  10 & -14.8  &  122    &        & \\
20010903.179      &  2.0m Coude   &  52156.6006  &   8 & -15.0  &  120    &        & \\
20011220.0011     &  Elodie       &  52264.3216  &  30 & -11.4  &         &      & \\
20011221.0008     &  Elodie       &  52265.3375  &  60 & -11.7  &         &      & \\
20041117.0010     &  Elodie       &  53328.4980  &  33 & -16.2  & 102.1   & 168.3  &  \\
20120903.1        &  2.0m Coude   &  56173.5163  &  10 & -24.08 &  ---    &  \\  
20120903.2        &  2.0m Coude   &  56173.5235  &  10 & -24.41 &  ---    &  \\  
20130102.1        &  2.0m Coude   &  56295.3645  &  10 & -25.61 &   81    &  \\  
20130102.2        &  2.0m Coude   &  56295.3717  &  10 & -25.95 &   88    &  \\  
20130103.1        &  2.0m Coude   &  56296.3813  &  10 & -26.39 &   85.9  &  \\  
20130103.2        &  2.0m Coude   &  56296.3885  &  10 & -26.54 &   85.8  &  \\  
20141013.1        &  2.0m Coude   &  56944.4487  &  10 & -24.4  &  105.5  &  \\  
20141013.2        &  2.0m Coude   &  56944.4560  &  10 & -24.17 &  101.8  &  \\  
20141212.1        &  2.0m Coude   &  57004.4019  &  10 & -26.8  &   87.2  &  \\  
20141212.2        &  2.0m Coude   &  57004.4091  &  10 & -26.6  &  108    &  \\   
20151223.1        &  2.0m Echelle &  57380.4701  &  30 & satur. &  ---    &  144.0  & \\   
20151223.2        &  2.0m Echelle &  57380.4916  &  10 & -36.4  &  ---    &  146.7  & \\    
20151224          &  2.0m Echelle &  57381.3691  &  10 & -36.3  &  ---    &  145.3  & \\    
20151226          &  2.0m Echelle &  57383.3341  &  10 & -36.1  &  ---    &  145.1  & \\    
20151227          &  2.0m Echelle &  57384.3728  &   5 & -36.5  &  ---    &  145.2  & \\    
20160130          &  2.0m Echelle &  57418.2988  &   5 & -33.9  &  95.3   &  124.9  & \\    
20160923          &  2.0m Echelle &  57654.5749  &  10 & -30.8  &  ---    &  143.3  & \\    
20161211          &  2.0m Echelle &  57734.3811  &  20 & -32.7  &  63.0   &  133.5  & \\    
20170317          &  2.0m Echelle &  57830.2978  &  20 & -30.9  &  ---    &  126.1  & \\  
20171207.1        &  2.0m Echelle &  58095.2062  &   2 & -23.5  &  119.4  &  \\    
20171207.2        &  2.0m Echelle &  58095.2107  &  15 & -23.5  &  120.3  &  149.3  & \\    
20171207.3        &  2.0m Echelle &  58095.2226  &  15 & -23.4  &  119.6  &  150.1  & \\    
20171208.1        &  2.0m Echelle &  58096.203   &   2 & -22.7  &  117.8  &  \\     
20171208.2        &  2.0m Echelle &  58096.207   &  10 & -23.5  &  121.8  &  148.0  &  \\
20171220.2134     &  TIGRE        &  58108.6489  &   4 & -25.0  &  125.3  &  \\
20171221.1853     &  TIGRE        &  58109.5373  &   4 & -24.1  &  122.4  &  \\
20171222.1854     &  TIGRE        &  58110.5380  &   2 & -23.4  &  120.3  &  \\
20171223.1922     &  TIGRE        &  58111.5571  &   4 & -23.8  &  120.8  &  \\
20171224.1926     &  TIGRE        &  58112.5601  &   4 & -23.0  &  120.7  &  \\
20171230          &  2.0m Echelle &  58118.2314  &  20 & -21.9  &  125.7  &  148.7  & \\
20180101          &  2.0m Echelle &  58120.230   &  20 & -19.8  &  121.5  &  143.7  & \\
20180103.1928     &  TIGRE        &  58122.5614  &  15 & -20.5  &  122.2  &  \\
20180103.2156     &  TIGRE        &  58122.6638  &  15 & -21.2  &  123.9  &  \\

\hline                                                           
  \end{tabular}                                                  
  \label{tab.obs}
\end{table*}

\addtocounter{table}{-1} 

\begin{table*}
\caption{Continued.}             
\centering
\begin{tabular}{lclrr ccr ccc  cccccr} 
\hline
20180104.0017     &  TIGRE        &  58122.7622  &  15 & -20.2  &  124.1  &  \\
20180104.1921     &  TIGRE        &  58123.5563  &  15 & -21.3  &  123.8  &  \\
20180105.1959     &  TIGRE        &  58124.5832  &  15 & -21.1  &  121.9  &  \\
20180105.2219     &  TIGRE        &  58124.6805  &  15 & -21.3  &  121.8  &  \\
20180106.0039     &  TIGRE        &  58124.7775  &   5 & -20.7  &  121.3  &  \\
20180106.1923     &  TIGRE        &  58125.5582  &  15 & -20.7  &  121.1  &  \\
20180106.2149     &  TIGRE        &  58125.6594  &  15 & -21.0  &  121.0  &  \\ 
20180107.0014     &  TIGRE        &  58125.7601  &  15 & -20.8  &  122.1  &  \\
20180126.01       &  2.0m Echelle &  58145.2150  &   2 & -18.34 &  106.3  &  \\       
20180126.02       &  2.0m Echelle &  58145.2170  &   3 & -18.44 &  106.1  &  \\      
20180126.03       &  2.0m Echelle &  58145.2201  &  15 & -18.05 &  106.8  &  \\      
20180126.04       &  2.0m Echelle &  58145.2312  &  20 & -18.64 &  108.5  &  141.3  & \\      
20180201.01       &  2.0m Echelle &  58151.2218  &  20 & -18.3  &  122.4  &  141.6  & \\
20180201.02       &  2.0m Echelle &  58151.2312  &   5 & -16.7  &  122.5  &  \\
20180402.01       &  2.0m Echelle &  58211.2832  &  10 & -14.7  &  105.8  &  115.0  & \\  
20180402.02       &  2.0m Echelle &  58211.2879  &   2 & -14.5  &  103.6  &  \\
20180802          &  TIGRE        &  58332.9723  &  10 & -17.83 & \\
20180818          &  TIGRE        &  58348.9210  &  10 & -18.89 & \\
20180905          &  TIGRE        &  58366.8712  &  10 & -19.62 & \\
20180925          &  TIGRE        &  58386.8679  &  10 & -20.61 & \\
20181010          &  TIGRE        &  58401.8028  &  10 & -21.51 & \\
\hline                                                           
  \end{tabular}                                                  
  \label{tab.obs}
\end{table*}

\end{document}